\documentstyle[a4,12pt]{article}

\begin{document}
 
\title{{\bf CONSIDERATIONS IN THE TIME-ENERGY UNCERTAINTY RELATION FROM THE VIEWPOINT OF HYPOTHESIS TESTING}}

\author{Keiji Matsumoto\\ National Institute of Informatics\\ 2-1-2 Hitotsubashi, Chiyoda-ku, Tokyo 101-8430, Japan\\ {e-mail: keiji@nii.ac.jp} \vspace{2em}\\ 
 Susumu Osawa \\  Faculty of Mathematics, Kyushu University\\ 744 Motooka, Nishi-ku, Fukuoka-city, Fukuoka 819-0385, Japan\\ {e-mail: s-osawa@math.kyushu-u.ac.jp}} 
\date{}
\maketitle

\renewcommand{\thefootnote}{}
\footnote[0]{This paper had appeared in Quantum Communication, Computing, and Measurement, Plenum Press, (1997) and is based on the talk given at the Quantum Communication Meeting held in 1996.}
\renewcommand{\thefootnote}{\arabic{footnote}}
\section{INTRODUCTION}
\quad The purpose of this study is to investigate time-energy uncertainty relation from the viewpoint of hypothesis testing.

  There are various derivations of time-energy uncertainty relation, and interpretation of $\Delta t$ is also various. The most acceptable derivation is that the relation is derived from the condition that the state of a system can hardly be distinguished from the initial state. For example, it is derived in the explanation  of the sudden approximation in Messiah $\cite{Messiah}$.   The outline is as follows.

\vspace{1cm} 
We suppose the Hamiltonian to change-over in a continuous way from a certain initial time $t_0$ to a certain final time $t_1$. We put
 \begin{equation}
    \Delta t = t_1-t_0
\end{equation}
  and denote by $H(t)$ the value taken by the Hamiltonian at time $t$. 

 Let $|0 \rangle $ denote the state vector of the system at time $t_0$ , $Q_0$ the projector onto the space of the vectors orthogonal to $|0\rangle$, and $U(t_1, t_0)$ the time evolution operator from $t_0$ to $t_1$. Supposing $|0\rangle$ to be of norm 1, we have 
\begin{equation}
 Q_0 = 1- |0\rangle  \langle 0 |.
\end{equation}
 
  The sudden approximation consists in writing 
\begin{equation}
U(t_1,t_0)|0\rangle \approx |0\rangle .
\end{equation}
Messiah regarded a probability $w$ as that of finding the system in a state other than the initial state and interpreted it to be  a measure of the error involved in this approximation:
\begin{equation}
   w=  \langle 0 | U^ {\dag}(t_1, t_0) Q_0 U(t_1, t_0)|0 \rangle \label{eq:w}.
\end{equation}
 
One obtains the expansion of $w$ in powers of $\Delta t$ by the perturbation method. Put 

\begin{equation}
   \overline {H} = \frac{1}{ \Delta t} \int_{t_0}^{t_1} H(t) dt.
\end{equation}

We then have 
\begin{equation}
 w= \frac {\Delta t^2}{\hbar^2}\langle 0 |\overline{H}Q_0 \overline{H} |0 \rangle + O(T^3). 
\end{equation} 
And since 
\begin{equation}
\langle 0 |\overline{H}Q_0 \overline{H} |0 \rangle =\langle 0|\overline{H}^2|0 \rangle - \langle 0|\overline{H}|0 \rangle ^2=(\Delta \overline{H})^2
\end{equation}
where $\Delta \overline{H}$ is the root mean squre deviation of the ovservable $\overline{H}$ in the state $|0\rangle$, one has 
\begin{equation}
 w= \frac{\Delta t^2 (\Delta \overline{H})^2 }{\hbar^2}+ O(T^3).
\end{equation}
 Thus the condition for the validity of the sudden approximation, $w\ll1$, requires that 
\begin{equation}
 \Delta t \ll \frac{\hbar}{\Delta \overline{H}}
\end{equation}
 
\vspace{1cm}
We can point out some  questions about the derivation of the relation. Messiah remarked  that $w$ is ``the probability of finding the system in a state other than the initial state'' and  the condition that the state of a system can hardly be distinguished from the initial state is $w \ll 1$. The first question is that the physical meaning of ``finding the system in a state other than the initial state'' is so ambiguous that the above condition cannot have a firm basis. We can find the state of the system only through measurements. Therefore, the degree of discernibility between the two states is dependent on the way of detection of the system.  The second question is that the detection scheme is not shown in Messiah's discussion and the indicator of discernibility is not shown from this point of view.    

 In this study, we investigate these questions from the viewpoint of hypothesis testing. 

\section{TIME-ENERGY UNCERTAINTY RELATION FROM \newline THE VIEWPOINT OF HYPOTHESIS TESTING}
\subsection{Appropriate Indicator of Discernibility}
\quad We investigate pure state in the following discussion.  The scheme of detection of the system should be constructed from a viewpoint of measurement and the decision rule of measurement outcomes.  Here, we propose an appropriate indicator of discernibility by constructing the best detection scheme. 
Put $n$ copies of state $\rho_t$, where $t$ is a time parameter. Consider the following hypothesis teting problem about a parameter $t$.
\begin{eqnarray*} \label{quantum test}
 & &         H_0: \quad \rho_ t=\rho_{t_0}\quad \quad \mbox{( null hypothesis)} \\
 & &    	     H_1:  \quad \rho_t=\rho_{t_1} \quad \quad\mbox{ (alternative hypothesis)}   
\end{eqnarray*}
  From hypothesis testing theory, the power of this test could represent discernibility between the states. Therefore, we define an indicator of discernibility between $\rho_{t_0}$ and $  \rho_{t_1}$ as a maximum power of test.    
 Then let us construct the test that maximizes the power of test $\gamma$. Since the probability distribution of measured value is determined  by parameter $t$ and measurement  $M$, two steps is needed to maximize $\gamma$ in the test. The first step is to select the most powerful test based on Neyman-Pearson's theorem subject to a fixed measurement. The second step is to select measurement in order to maximize $\gamma$ of the most powerful test dependent on measurement. These processes are  called optimization of the test. The selected test and measurement by optimization are called optimum test and optimum measurement respectively.  Thus, the indicator of discernibility  is the power of the optimum test.  

\subsection{Asymptotic Behavior of The maximum Power of Test}
\quad Let us consider the power of test and the optimum measurement when $\Delta t=t_1-t_0$ is very small and $n$ is very large. 

 To begin with, consider the first step. From stein's lemma ( see Appendix), the maximum power of test subject to a fixed measurement $M$ is 
\begin{equation}
  \gamma_M  \approx 1- \exp[-nD(p_{t_0} \| p_{t_1})],  \label{gamma of kullback}
\end{equation}
where $D(p_{t_0} \| p_{t_1})$ is Kullback divergence defined by $(\ref{kullback})$ in appendix, $p_{t_0}$ and $p_{t_1}$ probability distriibution of measured value at time $t_0$ and $t_1$.  
Because of $(\ref{gamma of kullback})$ and $(\ref{kullback and fisher})$, the power of test is written as 
\begin{equation}
\gamma_M \approx 1-\exp[-\frac{n}{2}J_M(t_0)(\Delta t)^2]+o((\Delta t)^2)\qquad(\Delta t \ll 1),
\label{kensyuturyoku}
\end{equation}
where $J_M(t_0)$ is classical Fisher information for the classical model $p(x|t_0)= {\rm Tr}\rho_{t_0}M(x)$ with a measurement $M$ defined as follows:
\begin{equation}
J_M(t_0) \stackrel{\rm def}{=} \lim_{t \rightarrow t_0} \Sigma_x \frac{\dot p(x|t)^2}{ p(x|t)}.
\end{equation}
  Then  consider the second step. We select the measurement which maximize $ \gamma_M$. Because of  $(\ref{kensyuturyoku})$, the optimum measurement maximizes  classical Fisher information $J_M(t_0)$. From the relation between classical  and quantum Fisher information $(\ref{quantum and classical fisher})$ in appendix, the optimum measurement $M_{opt}$ is one which satisfies
\begin{equation}
J_{M_{opt}}(t_0)= J^s(t_0), 
\end{equation}
where $J^s(t_0)$ is quantum Fisher information defined as follows:
\begin{equation}
J^s(t_0) \stackrel{\rm def}{=} 4{\rm Tr}\rho_{t_0}(\frac{d \rho_{t_0}}{d t})^2. 
\end{equation}

According to the pure state quantum estimation theory \cite{Fujiwara}, we have 
\begin{equation}
 J^s(t_0)=  \frac{4}{\hbar^2}\Delta H^2.
\end{equation} 
Thus we have 
\begin{equation}
J_{M_{opt}}(t_0) =\frac{4}{\hbar^2}\Delta H^2.\label{JM} 
\end{equation}

 From $(\ref{kensyuturyoku})$ and $(\ref{JM})$, the  power of the optimum test is  
\begin{equation}
\gamma_{max}=1-\exp(-\frac{2n}{\hbar^2}\Delta t^2\Delta H^2)+o(\Delta t^2)\qquad(\Delta t \ll 1).
\end{equation}
If $\frac{2n}{\hbar^2}\Delta t^2\Delta H^2\ll1$ holds,
\begin{equation}
\gamma_{max} \approx \frac{2n}{\hbar^2}\Delta t^2\Delta H^2. 
\end{equation}

Now we can show the condition that $\rho_{t_1}$ can hardly be distingished from $\rho_{t_0}$  using $n$ data when $\Delta t \ll1$ and $n \gg 1$ are satisfied. As it  means $\gamma_{max} \ll 1$, we have
\begin{equation}
1-\exp(-\frac{2n}{\hbar^2}\Delta t^2\Delta H^2)+o(\Delta t^2) \ll 1 \qquad(\Delta t \ll 1),
\end{equation}
or
\begin{equation}
\frac{2n \Delta t^2\Delta H^2 }{\hbar^2}\ll 1.  
\end{equation}

\subsection{The Optimum Measurement}
\qquad  Denoting  by $\Pi$ the measurement which is made up of operators $Q_0$ and $1-Q_0$, we can easily prove that $\Pi$ is one of the optimum measurements as follows.  

By fixing a state $\rho_t$ and a measurement $\Pi$, measured value  follows the probability function $ p_i(t) \; (i=1,2)$:
\begin{eqnarray*}
 p_1(t)&=&{\rm Tr}[\rho_t(1-Q_0)],\\
 p_2(t)&=&{\rm Tr} [\rho_tQ_0]\\
        &=&1-p_1(t).
\end{eqnarray*}
Therefore, classical Fisher information is  
\begin{equation}
J_\Pi(t_0)= \lim_{t \to t_0} [\frac{\dot{p_1}(t)^2}{p_1(t)}+\frac{\dot{p_2}(t)^2}{p_2(t)}].
\end{equation}
This limit is intermediate form, but $p_1(t)$ is easily expanded as follows:
\begin{eqnarray*}
p_1(t)&=&1+\dot{p_1}(t_0)(t-t_0)+\frac{1}{2}\ddot{p_1}(t_0)(t-t_0)^2+ \cdots\\
      &=&1-\frac{1}{\hbar^2}[\langle 0|H^2|0\rangle-(\langle 0|H|0\rangle)^2](t-t_0)^2+\cdots.
\end{eqnarray*}  
Hence, 
\begin{eqnarray}
J_{\Pi}(t_0)&=& \lim_{t\to t_0} \frac{(\dot{-p_1}(t))^2}{1-p_1(t)}\nonumber \\
          &=&-2\ddot{p_1}(t_0) \nonumber \\
          &=&\frac{4}{\hbar^2}\Delta H^2. \label{JP}
\end{eqnarray}
From $(\ref{JM})$ and $(\ref{JP})$, $\Pi$ is one of the optimum measurements.  A probability $w$ is that of a measured value of this measurement which supports $H_1$. 
\vspace{1em}
\hspace{-18pt}
\section{\Large\bf CONCLUSION AND DISCUSSION}
\vspace{1em}
A maximum power of test in the hypothesis testing $H_0: \; \rho_ t=\rho_{t_0}\quad        H_1:  \; \rho_t=\rho_{t_1}$ can be regarded  as an indicator of discernibility between the states.         
The condition that $\rho_{t_1}$ can hardly be distinguished from $\rho_{t_0}$  using $n$ data is 
\[
1-\exp(-\frac{2n}{\hbar^2}\Delta t^2\Delta H^2)+o(\Delta t^2)\ll 1 \qquad(\Delta t \ll 1), 
\]
or if $ \frac{2n}{\hbar^2}\Delta t^2\Delta H^2\ll1$,
\[
\frac{2n \Delta t^2\Delta H^2 }{\hbar^2}\ll 1.
\]
This condition represensts time-energy uncertainty relation from the viewpoint of hypothesis testing.
Measurement $\Pi$  made up of opetators $Q_0$ and $1-Q_0$ is one of the optimum measurements. A probability  $w$ is that of a measured value of this measurement which supports $H_1$.  It is remarcable that the previous study has suggested the optimum measurement that maximizes the power of test.   

\vspace{1em}
\hspace{-18pt}{\Large\bf Appendix }
\vspace{1em}

\quad Here we give a brief summary of the conventional hypothesis teting theory and related fields. 

  Suppose that random variables $X_i \;(i=1 \cdots, n)$ obey the probability distribution $p(x|\theta)$ with a given  parameter $\theta \in\Theta\subset {\rm {\bf R}}$ . Simple hypothesis testing about parameter    $\theta$ is as follows:
\begin{eqnarray*} 
 & &         H_0: \quad \theta=\theta_0\quad \quad \mbox{( null hypothesis)} \\
 & &        H_1:  \quad \theta=\theta_1 \quad \quad\mbox{ (alternative hypothesis)}   
\end{eqnarray*}
  We consider  nonrandomized test  based  on $n$ data.  Random variables $X_1,X_2, \cdots ,X_n$ are independent  and obey identical probability distribution $p(x|\theta)$.  $(X_1,X_2, \cdots , X_n)$ is denoted by $X$.  A hypothesis testing rule is a partition of the measurement space into two disjoint sets $U_0$ and $U_1=U_0^c$ . If observation value $x$ is an element of $U_0$, we decide that $H_0$ is true; if $x$ is an element of $U_1$, we decide $H_1$ is true. 

  Accepting hypothesis $H_1$ when $H_0$ actually is true is called a type I error, and the probability of this event is denoted by $\alpha$. Accepting hypothesis $H_0$ when $H_1$ actually is true is called a type II error, and the probability of this event is denoted by $\beta$. 

   The problem is to specify $(U_0, U_1)$ so that $\alpha$ and $\beta$ are as small as possible. This is not yet a well-defined problem because  $\alpha$ generally can be made smaller by reducing $U_1$, although $\beta$ thereby increases. The Neyman-Pearson point of view assumes that a maximum value of $\alpha$ given by $\alpha^*$ is specified and $(U_0, U_1)$ must be determined so as to minimize $\beta$ subject to the constraint that $\alpha$ is not larger than  $\alpha^*$.
We call $\gamma=1-\beta $ power of test, and the test with the maximum power of test subject to the above constraint is called the most powerful test.

A method for finding the optimum decision regions is given by the following theorem. 
\hspace{-18pt}{\bf Theorem ( Neyman-Pearson theorem)} 

Denote  joint density  function of random variables $X=(X_1, X_2, \cdots , X_n)$ by 
\[
 p_n(x|\theta)= \Pi_{i=1}^n p(x_i|\theta), \quad x=(x_1,x_2,\cdots x_n),
\]
 and put
\begin{equation}
 \Lambda_n \equiv \frac{ p_n(x|\theta_1)}{ p_n(x|\theta_0)}.
\end{equation} 
When a constant $k$ is set so that  
\[
 \int_{-\infty}^{\infty} \cdots   \int_{-\infty}^{\infty} \phi^*(x)p_n(x)dx  
 = \alpha^*
\]
holds, the regions of the most powerful test are determined as
\begin{eqnarray*}
& & U_0= \{ x:  \quad \Lambda_n \le k \} \\
& & U_1 =\{ x: \quad  \Lambda_n > k \},
\end{eqnarray*}
where $\phi^*(x)$ is  the function which is defined as
\[
\phi^*(x) = \left\{
          \begin{array}{@{\,}ll}
            1 & (\Lambda >k)\\ 
            0 & (\Lambda \le k).
          \end{array}
	\right.  
\]

\vspace{1cm}
The asymptotic behavior can be described in the following lemma. 

\hspace{-18pt}{\bf Theorem (Stein's lemma)}

Let $\alpha^* \in (0,1)$be given. Suppose that observaton  consists of n independent measurements. Let $\beta^*$ be the smallest probability of type II error over all decision rules such that the probability of type I does not exceed $\alpha^*$. Then all $\alpha^* \in (0,1)$, 
\begin{equation}
    \lim _{n \to \infty }(\beta_n^*)^{\frac{1}{n}}= \exp[-D(p_{\theta_0}\|p_{\theta_1})]. 
\end{equation}
Here,  $ D(p \| q)$ is called Kullback divergence and defined  as 
\begin{equation}
D(p \| q) \stackrel{\rm def}{=} E_p[ \log \frac{q}{p}], \label{kullback} 
\end{equation}
where $p$ and $q$ are probability distributions and  $E_p$ means expectation by $p$.

\vspace{1cm}
On the other hand, the following relation between Fisher information in classical information theory ( we call it classical Fisher information)  and Kullback divergence holds$(\cite{Nagaoka})$
\begin{equation} 
 D(p_{\theta+\Delta \theta}\| p_\theta)=\frac{1}{2}J(\theta)(\Delta \theta)^2+o((\Delta \theta)^2), 
\label{kullback and fisher}
\end{equation}
where $J(\theta)$ is classical Fisher information for the classical model $p_{\theta}$. 


Generally, the maximum value of classical Fisher information of a given state $\rho_{\theta}$ equals quantum Fisher information  \cite{Nagaoka87}:
\begin{equation}
     J^s(\theta)=\max_M J_M(\theta), \label{quantum and classical fisher}
\end{equation}    
 where $J^s(\theta)$ is quantum  Fisher information and $J_M(\theta)$ is classical Fisher information for the classical model $p(x|\theta)= {\rm Tr}[\rho_{\theta}M(x)]$ with a measurement $M$.


\begin{thebibliography}{99}


\bibitem{Fujiwara}
      Fujiwara, A. and H. Nagaoka, ``Quantum Fisher metric and estimation for pure state models,'' Phys. lett, 201A, 119-124(1995).
\bibitem{Messiah}
       Messiah, A.,
        ``MECANIQUE QUANTIQUE,'' 
	 Dunod, Paris 1959 
\bibitem{Nagaoka87}
     Nagaoka, H.,
      `` On Fisher information of quantum statistical models,'' SITA'87, 241-246,(1987).

\bibitem{Nagaoka}
     Nagaoka, H.,
     `` On the relation between Kullback divergence and Fisher information 
          - from calssical systems to quantum systems - ,''(1991).
\end{thebibliography}
\end{document}